\documentclass[
 aps, pra,
 amsmath,amssymb,
 11pt,
 final,
tightenlines,
 twoside,
 twocolumn,
 nofloats,
nofootinbib,
 superscriptaddress,
showkeys,
showkeywords,
 ]
{revtex4-2}

\usepackage[utf8]{inputenc}
\usepackage[english]{babel}
\usepackage{graphicx}
\usepackage{dcolumn}
\usepackage{bm}

\usepackage{ulem}

\input{maik.rty}

\setcitestyle{authoryear,round}
\def\squareforqed{\hbox{\rlap{$\sqcap$}$\sqcup$}}

\def\sq{\ifmmode\squareforqed\else{\unskip\nobreak\hfil
\penalty50\hskip1em\null\nobreak\hfil\squareforqed
\parfillskip=0pt\finalhyphendemerits=0\endgraf}\fi}

\def\utw{\smash{\rlap{\lower5pt\hbox{$\sim$}}}}

\def\udtw{\smash{\rlap{\lower6pt\hbox{$\approx$}}}}

\def\diameter{{\ifmmode\mathchoice
{\ooalign{\hfil\hbox{$\displaystyle/$}\hfil\crcr
{\hbox{$\displaystyle\mathchar"20D$}}}}
{\ooalign{\hfil\hbox{$\textstyle/$}\hfil\crcr
{\hbox{$\textstyle\mathchar"20D$}}}}
{\ooalign{\hfil\hbox{$\scriptstyle/$}\hfil\crcr
{\hbox{$\scriptstyle\mathchar"20D$}}}}
{\ooalign{\hfil\hbox{$\scriptscriptstyle/$}\hfil\crcr
{\hbox{$\scriptscriptstyle\mathchar"20D$}}}}
\else{\ooalign{\hfil/\hfil\crcr\mathhexbox20D}}%
\fi}}

\def\be{\begin{equation}}
\def\ee{\end{equation}}
\def\ba{\begin{eqnarray}}
\def\ea{\end{eqnarray}}

\def\ltsima{$\; \buildrel < \over \sim \;$}
\def\simlt{\lower.5ex\hbox{\ltsima}}
\def\gtsima{$\; \buildrel > \over \sim \;$}
\def\simgt{\lower.5ex\hbox{\gtsima}}

\usepackage[dvipsnames]{color}
\begin{document}

\selectlanguage{English}



\title{The influence of dust destruction on gas cooling}

\author{\firstname{S.~A.}~\surname{Drozdov}}
 \email{sai.drozdov@gmail.com}
 \affiliation{P.N. Lebedev Physical Institute of the Russian Academy of Sciences, Moscow, Russia}

\author{\firstname{M.~P.}~\surname{Yudkevich}}
 \affiliation{P.N. Lebedev Physical Institute of the Russian Academy of Sciences, Moscow, Russia}
 \affiliation{M.V. Lomonosov Moscow State University, Faculty of Space Research, Moscow, Russia}

\begin{abstract}
The observed dust abundance in the early Universe significantly exceeds the predictions of models assuming its efficient destruction at supernova shock wave fronts. We investigate the effect of dust on gas cooling behind the shock wave front, taking into account both dust cooling and thermal sputtering of dust grains for various interstellar dust models. Isochoric cooling of a gas element is considered for various initial temperatures (from $3\cdot 10^{5}$~K to $3\cdot 10^{7}$~K) and metallicities (from $10^{-2}$~Z$_{\odot}$ to 2~Z$_{\odot}$). It is shown that at metallicities below 0.1 Z$_{\odot}$, the effect of dust on gas cooling is negligible. The abundance of small grains in some models significantly accelerates cooling at temperatures below $10^{6}$~K, whereas large grains dominate cooling at high temperatures ($> 3\cdot 10^{6}$~K). For an initial temperature $T_{g,0} > 3\cdot 10^{6}$~K, less than 10\% of the dust mass survives for some models of the initial size distribution. The maximum survival rate (up to 20\% at $T_{g,0} = 3\cdot 10^{6}$~K and solar metallicity) is achieved for the model with the shallowest grain size distribution. The role of gas inhomogeneities is discussed: dust stripping from clouds by a shock wave can both decrease the cloud lifetime and contribute to the creation of dust tails, where the cooling of hot gas is accelerated, thereby increasing the dust survival rate.
\end{abstract}


\maketitle

\section{INTRODUCTION}

Observations made in recent decades have shown a discrepancy between the observed amount of dust in galaxies and the assumed rate of its destruction, both in the local Universe and at high redshifts ($z > 5$) (\cite{Shchekinov&Nath25}).

According to current data (\cite{Schneider24}), the main source of interstellar dust formation in the early Universe is supernova explosions, where dust is formed in a gas highly enriched with metals ejected from the exploded star.

However, supernova (SN) explosions also actively destroy dust (\cite{Draine79}) that gets behind the shock front, where the gas is strongly heated. In such a hot environment with temperatures above $10^{6}$~K, dust begins to efficiently undergo thermal sputtering (\cite{Draine2011}) on a timescale of $\tau \sim 10^{5}(1 + T_{6}^{-3})a_{0.1}n_{g}$~years, where $T_{6} = T_{g}/10^{6}$~K, $T_{g}$ is the gas temperature, $n_{g}$ is the gas number density, $a_{0.1} = a/0.1$~$\mu$m and $a$ is the grain size.

At the same time, bright and ultraluminous infrared galaxies are observed at redshifts up to $z\sim 8$. Their high infrared luminosity is explained by a large abundance of dust in the medium up to $M_{d}/M_{star}\sim 10^{-3} -10^{-2}$, where $M_{d} \sim 10^{8}M_{\odot}$ is the dust mass and $M_{star}$ is the stellar mass of the galaxy (\cite{Akins23}). In addition, there is observational evidence that at a redshift of $z\sim14$ there is already a noticeable dust mass $M_{d}\backsimeq 5\cdot 10^{4}M_{\odot}$ (\cite{Carniani24}). Such an observed dust abundance is difficult to explain if one assumes that the fraction of dust surviving after a SN explosion is $\sim$1\% (\cite{McKee89}).

In works (\cite{McKee89, Jones94, Slavin15}), it is shown that in the Milky Way, shocks destroy dust at a rate of $\dot{M_{d}} \sim 0.05 M_{\odot}$ per year, which is greater than the rate of dust formation in our Galaxy ($\dot{M_{d}}\sim 10^{-3} M_{\odot}$ per year (\cite{Bocchio16})). However, numerous observations show that dust is widespread in the interstellar medium of our galaxy. Therefore, an important task is to search for a possible mechanism that can ensure the survival of dust in supernova remnants (SNR).

In recent works \cite{Vasiliev24, Scheffler26, Shchekinov25}, it was shown that taking into account inhomogeneities and dust cooling of gas leads to a significant fraction of surviving dust — up to 80\% of the initial mass of the ejected material. 
Being in dense and cold clouds, dust is not subjected to the aggressive environment, provided that the cloud survives until the SNR cools down.
In addition, taking into account gas cooling due to interaction with dust grains can lead to a decrease in cooling time, which should also affect destruction rates, allowing dust grains to survive in the gas behind the shock fronts. Gas cooling on dust was considered in works \cite{Dwek87, Smith96, MartGonz16}. 

In \cite{Smith96}, the analysis of the destruction effect on the gas cooling rate was carried out and the survival rate of dust in hot gas was estimated; however, gas cooling functions in collisional equilibrium (\cite{Raymond77, Sutherland93, Mazzotta93}) and a power-law dust distribution with a slope of $\alpha = 3.5$ from (\cite{Mathis77}) were used.

In this study, we consider gas cooling with destroying dust for various values of metallicity and initial gas temperature. 
We study the thermal behavior of a unit volume of gas with a non-equilibrium cooling function, while taking into account gas cooling on dust with various initial grain size distribution functions. 
Thus, this work is devoted to studying the effect of external conditions and various dust models on gas cooling behind the shock front and dust survival.

In the section "Model description", we discuss the model for calculating gas cooling and the effect of dust on this process. 
In addition, in this section, we describe the algorithm for calculating the evolution of the dust size distribution function. In the section "Results", we present the calculation results: the gas cooling time and the contribution of various dust grains to the cooling function. In the section "Discussion", we comment on the obtained results of gas cooling calculations and the effect of dust cooling on dust survival.

\section{Model description}
\subsection{Gas and dust cooling function}

The paper considers the isochoric cooling process of a gas volume element with dust after the passage of a shock. 
The thermal evolution of the heated gas will be determined using the equation:
\begin{equation}
    \frac{dT_{g}}{dt} = -\frac{2n_{g}}{3k_{B}}\Lambda(Z,T_{g})
\end{equation}
where $n_{g} = 4n_{0}T_{g,0}/T_{g}$, $n_{0}$ is the background gas density before the shock impact, $T_{g,0}$ is the initial gas temperature, $k_{B}$ is the Boltzmann constant, and $\Lambda (Z,T_{g})$ is the total cooling function (including gas cooling on dust), normalized to the gas number density.

In calculating the thermal evolution, we will take into account the absorption of energy from hot gas particles by dust grains:
\begin{equation}
\label{coolfunc}
    \Lambda (Z,T_{g}) = \Lambda_{g}(Z,T_{g}) + \Lambda_{d}(Z,T_{g})
\end{equation}
We consider non-equilibrium gas cooling functions $\Lambda_{g}(Z,T_{g})$, constructed in a tabular form for various metallicity values in the range from $Z/Z_{\odot} = 10^{-2}$ to $Z/Z_{\odot} = 2$, where $Z_{\odot}$ is the solar metallicity value, and gas temperatures in the range from $10$~K to $10^{8}$~K from the work \cite{Vasiliev2013}.

The dust cooling function is determined by the expression 
\begin{equation}
    \Lambda_{d}(Z,T_{g}) = n_{d}(\zeta)H_{coll}(n_{g},T_{g})/n_{g}^{2} 
\end{equation}
where $n_{d}(\zeta)$ is the dust number density in the gas, and $\zeta = 1/120\times Z/Z_{\odot}$ is the dust-to-gas mass ratio in the interstellar medium (in this work, it is assumed that this ratio is proportional to the gas metallicity). 
The function $H_{coll}(n_{g},T_{g})$ is the heat flux absorbed by dust as a result of collisions with gas particles. In this work, we use the expression for $H_{coll}(n_{g},T_{g})$ from \cite{Dwek87}:
\begin{multline}
    H_{coll}(n_{g},T_{g}) = \sum_{i}\int_{a_{min}}^{a_{max}}\left(\frac{32}{\pi m_{e}}\right)^{1/2}\pi a^{2}n_{g}\\
    \times (k_{B}T_{g})^{3/2}h_{i}(a,T_{g})\frac{dN(a)}{da}da
\end{multline}
Here, the summation is performed over different dust species, $\frac{dN(a)}{da}$ is the dust size distribution function, and the function $h_{i}(a,T_{g})$ is given by the expression:
\begin{equation}
    h_{i}(a,T_{g}) = 1/2\int x^{2}\xi_{i}(a,E)e^{-x}dx
\end{equation}
where $x = \frac{E}{k_{B}T_{g}}$, and $\xi_{i}(a,E)$ is a function describing the fraction of energy absorbed by a dust grain after the penetration of a gas particle, which depends on the dust species and particle types. Due to their high velocity, electrons collide with dust grains more frequently than other particles. Therefore, they are the main agent of heat transfer from gas to dust. For electrons, the energy absorption efficiency function of dust grains upon collision will be determined according to Eq. 4 from \cite{Dwek86}. We assume that the electron density is determined by the gas density $n_{e} = n_{g}$.

We set a grid of initial gas temperature values from $3\cdot 10^{5}$~K to $3\cdot 10^{7}$~K, which corresponds to shock velocities from $150$~km/s to $1500$~km/s, respectively, according to the Hugoniot relation (\cite{Zeldovich67}). 
The lower limit of the temperature range is determined by the fact that at lower values, dust cooling of the gas is no longer efficient; moreover, the dust grain destruction time becomes significantly longer than the characteristic gas cooling time. 
The upper limit of the temperature range is determined by the fact that the destruction of dust grains is too rapid, and they do not survive during the calculation.

As a result of the thermal evolution of the gas, dust can remain in conditions of high ambient gas temperatures for a long time, which causes the destruction of dust particles. This leads to a change in the cooling function ,both due to the direct change in gas temperature and due to the evolution of the dust grain size distribution function.

We solve the cooling equation using the 4th-order Runge-Kutta method with an automatic time step calculated based on the minimum dust grain lifetime. At each step, we estimate dust destruction and the evolution of the size distribution, which is taken into account in the gas cooling function. The algorithm for changing the dust size distribution is described in more detail below.
The calculation ends at a gas temperature of $10^{4}$~K. We assume that upon reaching a size of less than 10~\AA\, a dust grain is completely destroyed and does not affect the further thermal evolution of the gas. Additionally, the release of metals as a result of the evaporation of dust grain material and the influence of this process on the gas cooling function are neglected in this work.

\subsection{Dust destruction in hot gas}

Unlike heating, the ion flux, primarily protons, plays the main role in dust destruction. 
Due to the fact that we consider gas temperatures well above $10^{4}$~K, we assume the plasma to be fully ionized. The main destruction mechanism in our model is thermal sputtering associated with the random flux of incident protons. 
We neglect kinetic sputtering, as it was discussed in \cite{Vasiliev24} that the contribution from kinetic sputtering is about 15$\%$ of thermal sputtering. We also neglect collisions between dust grains and the evolution of the dust distribution function due to this process.

We set a grid of dust grain sizes uniformly distributed on a logarithmic scale from 10 \AA\, to 3000 \AA. The number of size bins in the distribution is 50. The dust consists of two species: silicates with a density of $3.2$~g/cm$^{3}$ and graphites with a density of $2.2$~g/cm$^{3}$ with an equal mass fraction between the species (\cite{Corrales16}).

The lifetime of a dust grain of radius $a$ surrounded by gas with temperature $T_{g}$ and density $n_{g}$ is determined by the expression (\cite{Polikarpova17}): 
\begin{equation}
    \tau(a,n_{g},T_{g})^{-1} = \frac{\mu_{i}m_{H}n_{g}v(T_{g})}{a\rho_{i}}Y(T_{g})
\end{equation}

\begin{equation}
\label{taulife}
    \tau^{-1} = \frac{2.4\cdot 10^{-19}\mu_{i}\eta_{i}}{\rho_{i}}\frac{n_{g}}{a}T_{6}^{1/2}e^{-0.54/T_{6}}, \, \text{s}^{-1}
\end{equation}
where $T_{6} = T_{g}/10^{6} K$, $\rho_{i}$ is the dust material density, $i$ is the dust grain species index, $\mu_{i}$ is the molecular weight of particles leaving the dust grain surface as a result of sputtering ($\mu_{C}$ = 12 and $\mu_{Si}$ = 28).
The function $Y(T_{g}) = 10^{-2}\eta_{i}e^{-0.54/T_{6}}$, where $\eta_{C}$ = 2.5 and $\eta_{Si}$ = 7.5 from \cite{Polikarpova17}.

During the destruction process, a dust grain will gradually transition to a bin of a smaller size. 
Thus, we can write that the number of dust grains in the $j$-th bin as a function of time will be determined by the expression:
\begin{equation}
\label{evoldstrb}
    N_{j}(t + \delta t) = N_{j}(t) - \delta N_{j}(t) + \delta N_{j + 1}(t)
\end{equation}
where $\delta N_{j}(t)$ is the number of dust grains that, during the time interval $\delta t$, were destroyed and transitioned to a smaller size bin with index $j-1$, and $\delta N_{j + 1}(t)$ is the number of destroyed dust grains from the larger size bin $j+1$ that transitioned to the size bin with index $j$. 
The change in the number of dust grains over a small time interval $\delta t$ can be defined as:
\begin{equation}
\label{dNdt}
    \delta N_{j}(t) = \frac{\delta tN_{j}(t)}{\tau(a_{j},n_{g},T_{g})}
\end{equation}
Thus, we can rewrite expression \ref{evoldstrb} using expressions \ref{taulife} and \ref{dNdt}.
\begin{equation}
    N_{j}(t + \delta t) = N_{j}(t) - \frac{n_{g}\delta t}{\psi(T_{g})}\left(\frac{N_{j}(t)}{a_{j}} - \frac{N_{j+1}(t)}{a_{j+1}}\right)
\end{equation}
where the function is: $\psi (T) = 2.4\cdot 10^{-19}\mu_{i}\eta_{i}/\rho_{i}T_{6}^{1/2}e^{-0.54/T_{6}}$.

In this work, we compare several models of the initial dust grain size distribution that are characteristic of different conditions of the interstellar medium through which shocks can pass.
\begin{itemize}
    \item A power law of the form $dN/da \propto a^{-\alpha}$ with a slope of $\alpha = 3.5$ (hereafter MRN) from \cite{Mathis77}. 
    \item A power law with a slope of $\alpha = 1.5$ (hereafter we will call it the distribution with $\alpha = 1.5$). Such a distribution is characteristic of galaxies at high redshifts (\cite{Maiolino04,Nozawa07}), where dust is injected into the interstellar gas as a result of SN explosions, and the number of small grains is small compared to large ones due to their destruction in regions of intense star formation
    (see, e.g., the discussion in section 2.2 of \cite{Nath25}). In addition, similar flat spectra of the dust grain size distribution are obtained in young galaxies as a result of numerical simulations (\cite{Nishida22}).
    \item The dust distribution model from \cite{Weingartner01} (hereafter WD), which takes into account the abundance of small carbonaceous grains $a\sim 10$~\AA.
    Distribution parameters: $b_{C} = 6\cdot 10^{-5}$ is the carbon abundance relative to hydrogen, and $R_{V} = 3.1$ is the extinction parameter. Such a distribution is characteristic of the interstellar medium with evolved dust in our Galaxy.
    \item A dust distribution characteristic of the ejected material after a SN explosion. We took the distribution from \cite{Sarangi15} (hereafter SC) for the time moment 2000 days after the explosion. The progenitor star mass is $15M_{\odot}$. This distribution corresponds to the dust ejected into the gas of a young remnant but has not yet experienced the impact of the reverse shock. This distribution is not preserved during the entire lifetime of the supernova remnant. By the time the reverse shock arrives, the dust in dense gas clouds will be distributed according to size, according to this model.
\end{itemize}
In this work, we compare the dust survival between these models for different values of the external conditions in which the dust is located, as well as the influence that these models have on the gas cooling rate.

\section{Results}

\begin{figure}[htbp]
   \centering
\includegraphics[width=1\columnwidth]{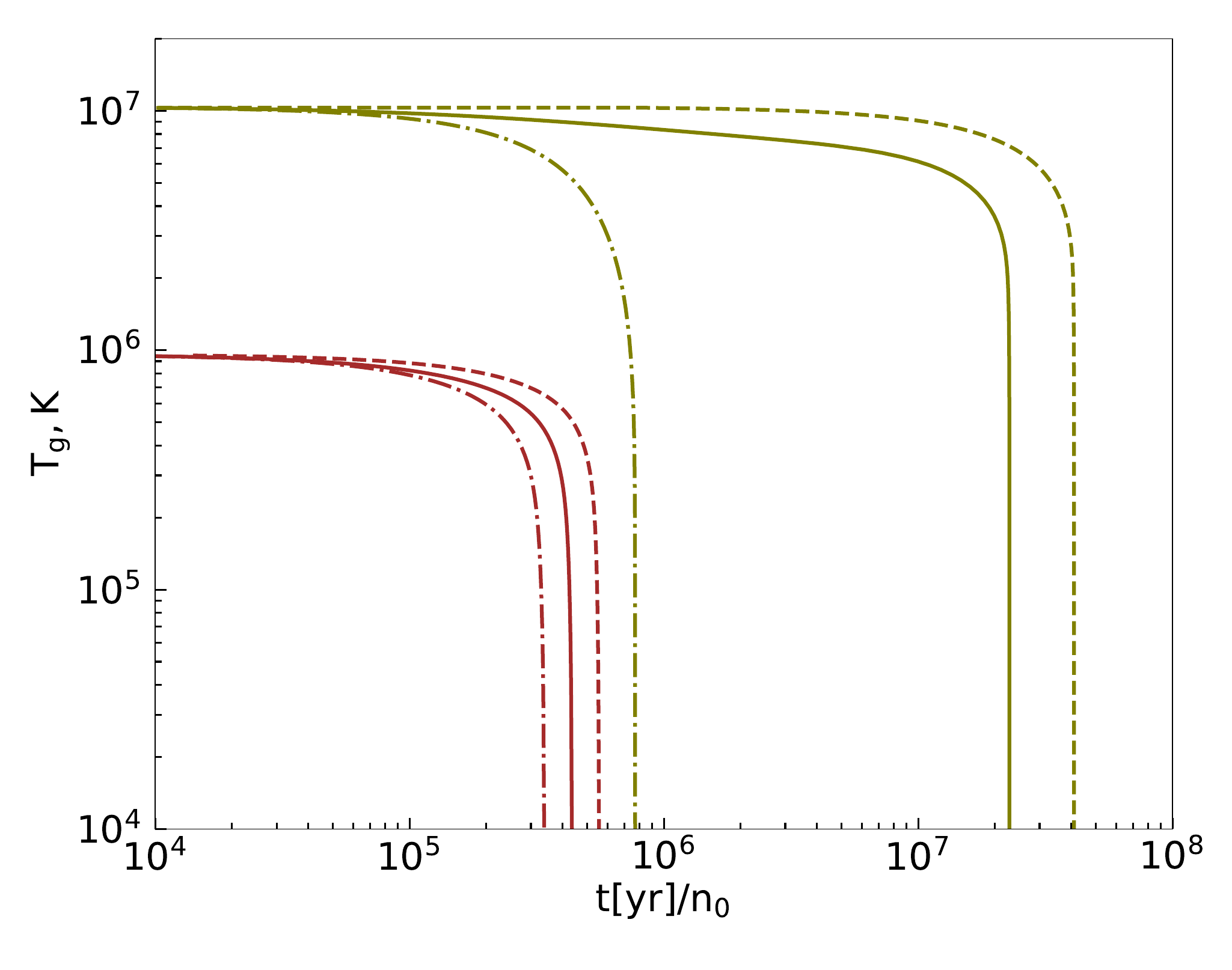}
   \caption{Temperature evolution of gas with various cooling regimes. Yellow lines show cooling for gas with an initial temperature of $10^{7}$~K, red lines show cooling for an initial temperature of $10^{6}$~K. Dashed lines show gas cooling without taking into account the effect of dust cooling, solid lines show cooling with dust and dust grain destruction, dash-dotted lines show cooling taking into account gas cooling on dust but without destruction. The dust is distributed according to the MRN power law. The gas metallicity is solar. The cooling time is normalized to the initial gas density $n_{0}$.
  }
   \label{Fig1}
\end{figure}

Fig. \ref{Fig1} shows the thermal evolution of a gas element with initial temperatures of $10^{6}$~K and $10^{7}$~K. It can be seen that taking into account gas energy losses due to interaction with dust significantly changes the gas cooling rate. In this example, the cooling time is reduced by almost a factor of two for an initial temperature of $10^{7}$~K when gas cooling on dust is considered. However, if we do not take into account dust destruction during cooling, the cooling becomes too intense, and the cooling time is reduced by almost a factor of 50 compared to the case of cooling without dust. 
For an initial temperature of $10^{6}$~K, the gas cooling time decreases by a factor of 1.3 when dust is added. If we do not take into account dust grain destruction, the gas cooling rate will increase by a factor of 1.6. This shows that when calculating the gas cooling rate, it is necessary to consider both the effect of dust on the cooling function and its destruction.

\begin{figure*}[!t]
   \centering
\includegraphics[width=\textwidth]{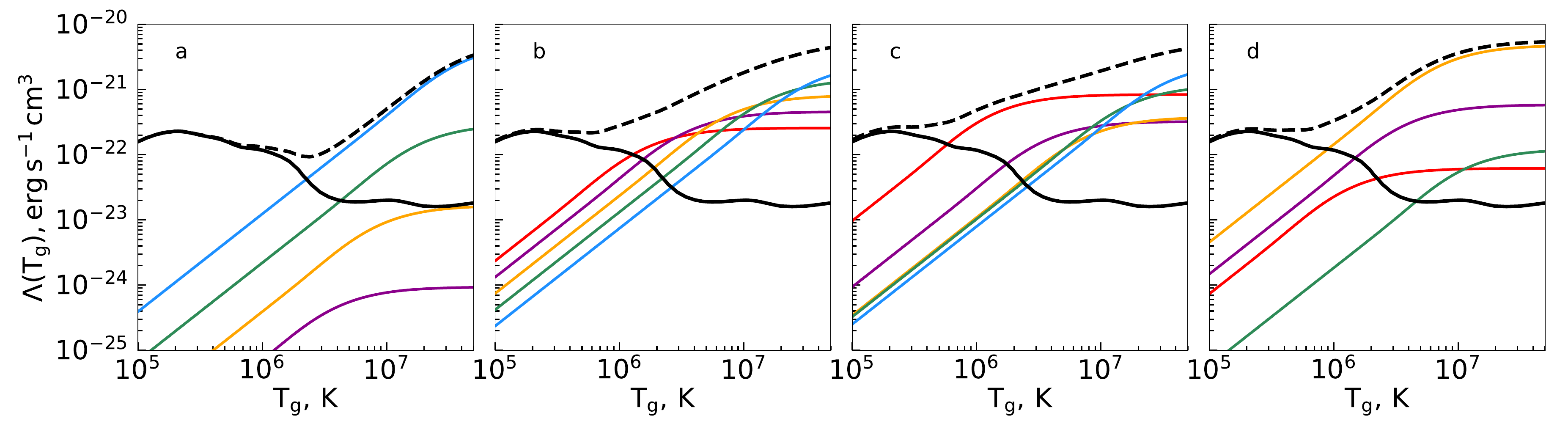}
   \caption{Gas and dust cooling functions for various dust grain size distributions. All panels consider solar metallicity. From left to right: a -- distribution with $\alpha =1.5$, b -- MRN distribution, c -- WD distribution, d -- SC distribution. The solid black curve on the panels is the contribution to the cooling function from gas, the black dashed line is the total cooling function with gas and dust. Colored lines show the contributions from dust in various size ranges: red line -- dust with sizes $a \in [10;30]$~\AA, purple -- with sizes $a \in (30;100]$~\AA, orange -- with sizes $a \in (100;300]$~\AA, green -- with sizes $a \in (300;1000]$~\AA\, and blue -- with sizes $a \in (1000;3000]$~\AA.  
  }
   \label{Fig2}
\end{figure*}

Dust affects the gas cooling function differently depending on the size distribution model and the dust grain radius, as can be seen in Fig. \ref{Fig2}. Thus, for the $\alpha = 1.5$ distribution (panel \textbf{a} in Fig. \ref{Fig2}), dust begins to affect gas cooling at $T_{g} > 2\cdot10^{6}$~K.
Large dust ($a > 0.1$~$\mu$m) constitutes the main fraction in this flat size distribution, so it makes the largest contribution to gas cooling.

The MRN and WD dust models (panels \textbf{b} and \textbf{c} in Fig. \ref{Fig2}, respectively) demonstrate that small grains ($a < 100$~\AA) have a significantly larger contribution to the $\Lambda(Z,T_{g})$ function. At the same time, in the WD model, this contribution is more noticeable than in the MRN distribution, becoming significant already at a gas temperature of $T_{g} \sim 3\cdot 10^{5}$~K and remaining so over a wide temperature range. At high gas temperatures $T_{g} > 10^{7}$~K, in the $\alpha = 1.5$, MRN, and WD models, the main sources of gas cooling are large grains $a > 300$~\AA.

In the SC dust model, the main contribution to gas cooling is made by grains with radii from 100 \AA\, to 300 \AA\, (see Fig. \ref{Fig3}, panel \textbf{d}). This is due to the fact that these grains constitute the main part of the SC distribution. 

\subsection{Evolution of the dust size distribution function}

\begin{figure*}[htbp]
   \centering
\includegraphics[width=\textwidth]{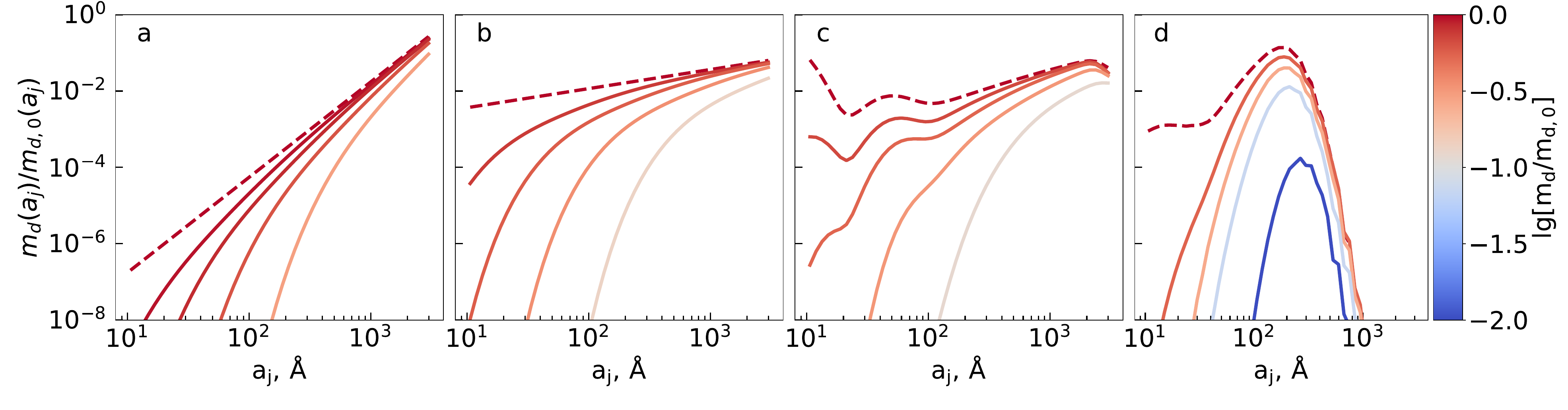}
   \caption{Evolution of dust grain mass distribution functions by sizes during gas cooling for different dust models. Initial gas temperature is $10^{7}$~K, metallicity is $Z/Z_{\odot} = 1$. The panels are arranged in the same way as in Fig. \ref{Fig2}, solid lines from top to bottom correspond to different time moments: $3\cdot10^{3}$, $10^{4}$, $3\cdot 10^{4}$, $10^{5}$ and $3\cdot 10^{5}$ years, the dashed line corresponds to the initial dust mass size distribution. The color shows the change in the total dust mass normalized to the initial dust mass.
  }
   \label{Fig3}
\end{figure*}

Fig. \ref{Fig3} shows the evolution of the dust mass distribution function as a function of time during the cooling of gas with an initial temperature of $T_{g,0} = 10^{7}$~K and metallicity $Z/Z_{\odot}=1$. 
The quantity $m_{d}(a_{j})/m_{d,0}$ is the mass of graphite and silicate dust particles contained in the size bin $a_{j}$, normalized to the dust mass before the onset of cooling.
It can be seen that small grains are destroyed first over time, as was expected. However, the gradual transfer of grains between sizes, when large ones move to a smaller size bin, prevents small grains from completely disappearing for a long time.

For all models, the fraction of dust mass that survived after gas cooling is less than 10\%, and for the SC model (panel \textbf{d}), this value is less than one percent. Dust survives best in the model with $\alpha = 1.5$ (panel \textbf{a}), which is explained by the larger contribution of large grains ($a > 500$~\AA) to the total dust mass, for which the destruction time under such conditions is longer than or comparable to the gas cooling time.

\subsection{Gas cooling time with dust}

\begin{figure*}[htbp]
   \centering
\includegraphics[width=\textwidth]{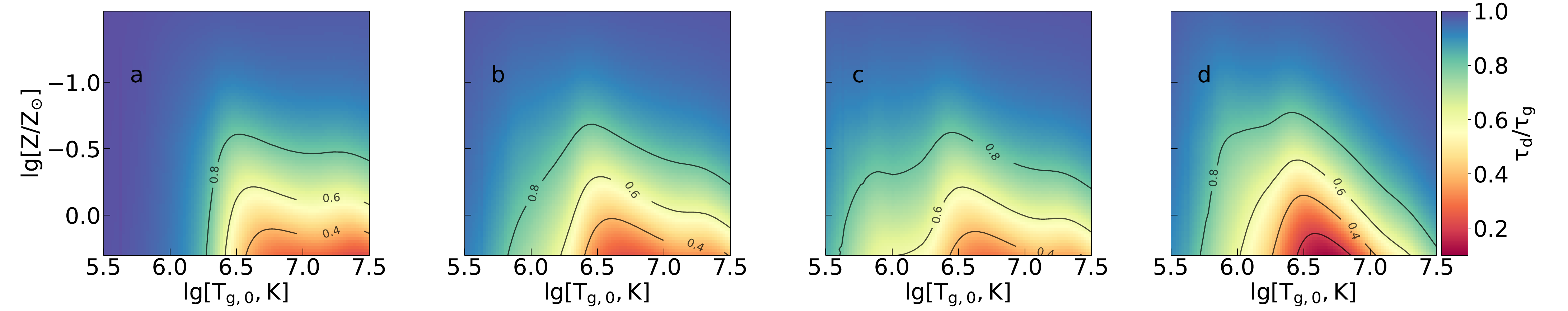}
   \caption{Ratio of the gas cooling time taking into account cooling on dust $\tau_{d}$ to the gas cooling time without dust $\tau_{g}$, constructed on a grid of gas parameters -- metallicity and initial gas temperature $T_{g,0}$. The four panels correspond to different dust models, as in Fig. \ref{Fig2}. The contour lines correspond to the cooling time ratio $\tau_{d}/\tau_{g}$: 0.8, 0.6, 0.4, and 0.2.
  }
   \label{Fig4}
\end{figure*}

For a grid of parameters (initial temperature and metallicity), we constructed the ratio of gas cooling times with and without taking into account cooling on dust. In Fig. \ref{Fig4}, one can see the cooling time ratios for different dust distribution models. It can be seen that the lower the metallicity, the smaller the contribution of dust to gas cooling, and the cooling time does not depend on the presence of dust. At a gas metallicity of $Z/Z_{\odot} < 0.1$, the effect of dust can be neglected. This is expected since the dust-to-gas mass ratio $\zeta$ is proportional to metallicity.

The dependence of the gas cooling rate on the initial temperature has a more complex character, and here the initial dust distribution is already important. Thus, for the dust distribution with $\alpha = 1.5$, it can be seen that at an initial gas temperature of $\lesssim 10^{6}$~K, the effect of dust is already imperceptible (Fig. \ref{Fig4}, panel \textbf{a}).

On the other hand, with the MRN distribution of grain sizes, as well as in the WD and SC models, where the number of small grains is large, the effect of dust is noticeable even at temperatures significantly below $10^{6}$~K. The effect of small grains on the cooling function is shown in Fig. \ref{Fig2}.  
At such values of the initial gas temperature, small dust does not have time to be noticeably destroyed and affects gas cooling even at low initial temperatures ($T_{g,0} < 10^{6}$~K).

If we consider high initial gas temperatures, then larger dust, which can survive longer in hot plasma with $T_{g} > 3\cdot 10^{6}$~K compared to small grains, will have an effect on cooling.

In the $\alpha = 1.5$ model, the main part of the dust mass is concentrated in large grains ($a \sim 1000$~\AA). They survive longer in a hot environment, cooling the gas, and have a strong effect on the evolution of the gas behind the shock front.

The behavior of dust in the MRN distribution and in the WD model is similar to each other. Even at the highest values of the initial temperature in these models, dust helps to cool the gas almost twice as fast at solar metallicity.

Newborn dust (SC model) is destroyed much faster during the thermal evolution of the gas than in the above models, due to the lower content of large grains. In addition, such dust makes a smaller contribution to gas cooling in cases of high initial temperature $ > 10^{7}$~K. However, at a metallicity of $Z/Z_{\odot} \approx 2$ and initial temperatures of $ \approx 3\cdot 10^{6}$~K, the gas cooling rate increases by more than 5 times due to the absorption of energy by dust grains.

\section{Discussion}

\begin{figure*}[htbp]
   \centering
\includegraphics[width=\textwidth]{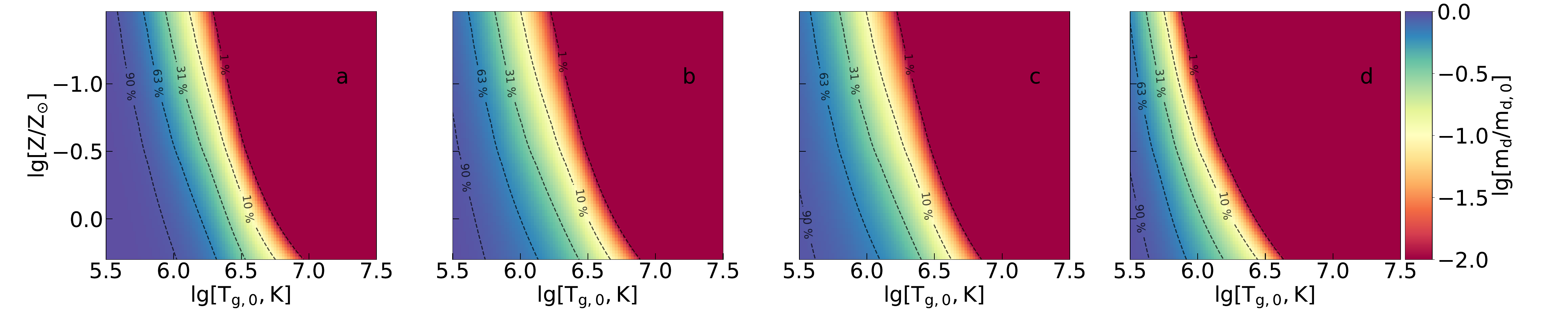}
   \caption{Ratio of the total dust mass remaining after gas cooling to the initial dust mass, for various gas parameters: metallicity and initial gas temperature. The four panels correspond to different dust models, as in Fig. \ref{Fig2}. The contour lines correspond to certain values of the percentage of surviving dust relative to the initial mass: 90 \%, 63\%, 31\%, 10\%, 1\%.
  }
   \label{Fig5}
\end{figure*}

The fraction of dust mass that survived after gas cooling is shown in Fig. \ref{Fig5}. For all distributions, it is clearly seen that above a certain temperature, the dust is almost completely destroyed. 
Dust in the model with $\alpha = 1.5$ survives best. For solar metallicity, the fraction of dust that survived gas cooling with an initial temperature of $3\cdot 10^{6}$~K is more than 20\%. In the MRN and WD models, under the same conditions, the fraction of surviving dust is $\sim$10\%.

This indicates that in the $\alpha = 1.5$ model, a larger fraction of dust grains can survive behind the shock fronts. Thus, in galaxies at high redshifts $z > 5$, where a flat power-law distribution is expected than in the local Universe (\cite{Maiolino04,Nozawa07}), dust more easily survives frequently passing shocks from SN explosions.

The WD distribution does not differ much from the MRN model in terms of dust survival.

Newborn dust from the SC distribution survives gas cooling worst, where a significant part of the mass is concentrated in grains of size $a\sim 300-500$~\AA . Such dust, apparently, can survive the impact of the reverse shock only in dense gas condensations, where the temperature can remain low for a long time (\cite{Scheffler26, Vasiliev24, Dedikov25}). Such dense gas clouds can appear due to thermal instabilities at early stages of the expansion of SNR in the ejected material of the progenitor star even before the formation of the reverse shock (\cite{Shchekinov25}).

For all dust models, it is clearly seen that with a decrease in the concentration of metals in the gas, and consequently in dust cooling, the survival rate of dust grains decreases significantly, which is associated with an increase in the cooling time. Thus, for the same initial temperature $T_{g,0} \approx 3\cdot10^{6}$~K, up to 30 $\%$ of the initial dust mass can be preserved in the case of solar metallicity (in the model with $\alpha = 1.5$), but for $Z/Z_{\odot} = 0.1$ the dust will be almost completely destroyed.

This work did not consider the processes of collisions between dust grains. This is primarily due to the simplified model: we solve a 0-dimensional problem of cooling of an elementary gas volume.
However, such an effect can influence the evolution of the dust grain distribution function and the gas cooling rate behind the shock fronts. 
In \cite{Tielens94}, this was considered in detail, and, as follows from it, it can be concluded that as a result of shattering, the number of large grains will decrease, while the number of small grains will increase. Such behavior will lead to the fact that, on the one hand, the gas cooling rate will increase, since there are more small grains, and on the other hand, this will lead to an increase in the dust destruction rate. 
Thus, the effect of dust shattering on gas cooling is an important task that should be considered in a separate study.

Gas cooling on dust helps a cloud surrounded by hot gas avoid evaporation for a longer time, which preserves the dust contained in it from destruction (\cite{MartGonz25}). Taking into account cooling on dust in the cloud reduces the ratio of the gas cooling time to the time it takes for the cloud to be destroyed by almost an order.

For $\zeta = 1/120$ and a cloud density of $n_{g}=10^{4}$~cm$^{-3}$, the cloud survives the cooling of gas heated by an shock with a velocity of $v_{sh}\sim 10^{3}$~km/s, which corresponds to a gas temperature behind the front of $T_{g} \sim 10^{7}$~K.

However, in \cite{Dedikov24} it was shown that an shock incident on a cloud can strip large dust ($a > 100$~\AA) from it, since the gas and the smallest dust ($a < 100$~\AA) are easily entrained behind the front. In their simulations, the effect is more pronounced the smaller the cloud size ($r < 0.3$~pc, where $r$ is the cloud radius). This means that a cloud from which all large dust has been stripped will be destroyed faster in hot gas, since dust cooling will be less efficient. 
For example, for a cloud of size $r = 0.03$~pc, all dust with a size of $a > 100$~\AA is stripped beyond its boundaries. If we assume that initially the dust in the cloud is distributed according to the MRN model, then after the shock impact, the dust-to-gas mass ratio $\zeta$ in it will become almost 8 times lower, which should reduce the time it takes for such a cloud to be destroyed in hot gas.

Such an effect can influence the estimate of the lifetime of clouds when solving problems of shock propagation through a turbulent medium.

\begin{figure}[htbp]
   \centering
\includegraphics[width=1\columnwidth]{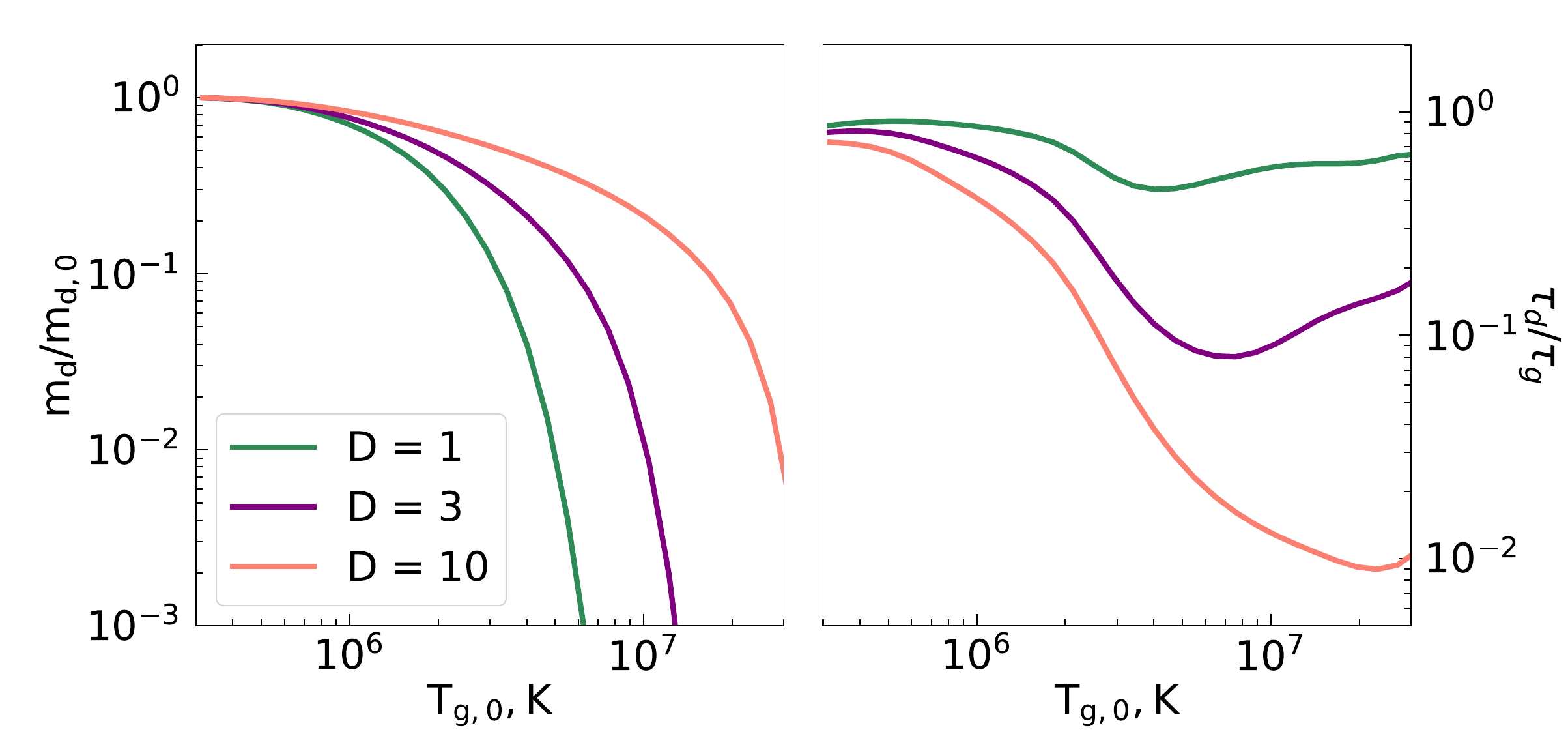}
   \caption{On the left panel -- the ratio of the total mass of surviving dust to the initial mass as a function of the initial gas temperature for three different values of dust abundance $D \equiv \zeta^{*}/\zeta$. On the right panel -- the ratios of the gas cooling time with dust to the gas cooling time without dust as a function of the initial gas temperature for three different values of dust abundance $D$. Gas metallicity $Z/Z_{\odot} = 1$, initial dust distribution -- MRN.
  }
   \label{Fig6}
\end{figure}

In \cite{Dedikov24}, the ratio between the density of the background medium and the cloud $\chi \equiv n_{cl}/n_{a} = 10$ was considered, where $n_{cl}$ is the gas density in the cloud, $n_{a}$ is the background density of the medium.
Dust stripped from the cloud by the incident shock can strongly change the gas cooling rate in the dust tail forming behind it. 
In the example above for a cloud of $r = 0.03$~pc, such stripped dust ends up in hot gas in a volume comparable to the size of the initial cloud, and noticeably increases the dust-to-gas mass ratio compared to the standard value $D \equiv \zeta^{*}/\zeta \sim \chi$, where $\zeta = 1/120$ is the initial ratio for $Z/Z_{\odot} = 1$, and $\zeta^{*}$ is the ratio of dust density to gas after the ejection of grains from the cloud plus background dust in the gas flow.
Such an increase in dust concentration will significantly affect the gas cooling rate, and similar dust tails behind evaporating clouds can accelerate gas cooling and contribute to dust survival behind the shock front. In Fig. \ref{Fig6}, it is clearly seen that with an increase in the dust-to-gas mass ratio, the fraction of dust that survives in the cooling gas noticeably grows. At $D = 10$ and an initial temperature of $T_{g,0} = 10^{7}$~K, about 20 \% of the initial dust mass survives after cooling, while at $D = 3$ -- 1 \% of the initial mass.
The effect of dust abundance $D$ is also noticeable on the gas cooling rate. For $D = 10$, the ratio $\tau_{d}/\tau_{g} \approx 10^{-2}$ in the case of an initial gas temperature of $T_{g,0} = 10^{7}$~K, and for $D = 3$, the cooling time ratio $\tau_{d}/\tau_{g} \approx  10^{-1}$.

The applicability of this result depends on the parameter $\chi$, since it is not entirely clear whether the process of dust stripping from the cloud will occur efficiently if the density ratio in the cloud and the background medium is higher. 
Conducting such numerical calculations will allow a better understanding of the processes of gas cooling behind the shock front in a cloudy medium and the rate of dust destruction in the interstellar gas.

\section{Conclusion}

Dust cooling is important to consider when studying the temperature evolution of hot gas. At the same time, thermal sputtering of dust in hot gas must also be taken into account, since it strongly affects the plasma cooling rate. 
In this work, we evaluated the effect of thermal sputtering processes of dust grains on the gas cooling rate behind the shock front, as well as the survival of dust in various models of dust grain size distribution.
As a result of the study, we obtained the following:
\begin{itemize}
     \item At a metallicity of $Z/Z_{\odot} < 0.1$, the effect of dust on gas cooling is negligible. Low metallicity promotes the destruction of dust grains due to the longer gas cooling time.
     \item The initial dust distribution function has a noticeable effect on the gas cooling rate. Small grains in the MRN and WD models significantly accelerate the cooling of gas with an initial temperature of $T_{g,0} < 10^{6}$~K.
     Gas cooling rates with high initial temperatures are significantly higher in the model with a flat dust distribution ($\alpha = 1.5$) compared to other models.
     This is due to the larger contribution of large grains to the total dust mass.
     \item At an initial gas temperature of $T_{g,0} > 3\cdot 10^{6}$~K, almost all dust is strongly destroyed in all models. Dust survives better in the model with a power-law spectrum slope of $\alpha = 1.5$.
     \item The stripping of dust from clouds under the impact of an incident shock can accelerate the cooling of hot gas in the dust tail behind the cloud, which will contribute to the survival of dust stripped from the cloud.
 \end{itemize}

\begin{acknowledgements}
The authors are grateful to E.O. Vasilyev for valuable remarks and discussions.
\end{acknowledgements}

\bibliographystyle{aspb1}
\bibliography{refs_dcool_arxiv.bib}

\end{document}